\documentclass[preprint,authoryear,12pt]{elsarticle}

\usepackage{amsmath,amsfonts,amssymb,amscd,amsthm,amsbsy}
\usepackage{bm}
\usepackage{mathabx}

\def\E{\Earth}

\usepackage{graphicx}
\graphicspath{{graphics/}}
\usepackage[font= footnotesize]{caption}
\usepackage[font=footnotesize]{subcaption}

\usepackage[table]{xcolor}
\usepackage{booktabs}
\usepackage{tabu}
\usepackage{threeparttable}
\usepackage{multirow}

\usepackage{algorithm}
\usepackage{algpseudocode}
\usepackage{tikz}
\usetikzlibrary{shapes.geometric, arrows.meta, positioning, shapes, arrows, calc}

\usepackage{fullpage}

\usepackage[%
breaklinks=true,%
colorlinks=true,%
pdfauthor={First Author et al.},%
pdftitle={Template for manuscripts in Advances in Space Research}%
]{hyperref}

\journal{Advances in Space Research}

\begin{document}

\begin{frontmatter}

\title{A dynamical coherency gate for state recovery: Statistical \\ requiem for the long arc of cislunar orbital mis-prediction}

\author{Binyamin Stivi}
\ead{bstivi@ucsd.edu}
\address{Department of Mechanical and Aerospace Engineering, UC San Diego, La Jolla, CA 92093, USA}

\author{Vishnuu Mallik}
\ead{vishnuu@planet.com}
\address{Planet Labs PBC, San Francisco, CA 94107, USA}

\author{Lamberto Dell'Elce}
\ead{lamberto.dell-elce@inria.fr}
\address{Universit{\'e} C{\^o}t{\'e} Azur and Inria Sophia Antipolis M{\'e}diterran{\'e}e, 06902 Valbonne, France}

\author{Aaron J. Rosengren\corref{cor}}
\cortext[cor]{Corresponding author}
\ead{arosengren@ucsd.edu}
\address{Department of Mechanical and Aerospace Engineering, UC San Diego, La Jolla, CA 92093, USA}

\begin{abstract}

We present a statistically grounded methodology for recovering physically consistent initial conditions from historical two-line element sets (TLEs), enabling accurate long-arc trajectory reconstruction for distant and highly eccentric Earth satellites. The approach combines numerical averaging of osculating orbital elements with Gaussian-mixture-model (GMM) filtering with robust fallback strategies to isolate a dynamically coherent subset of mean elements at a common reference epoch. From this filtered ensemble, a representative osculating element is reconstructed, yielding a recovered Cartesian state vector with predictive capability far exceeding that of raw TLE-based propagations and existing approaches. We apply this method to the case of OGO-1 (1964-054A), a spacecraft launched into a cislunar orbit and tracked intermittently over five decades. Despite large observational gaps and significant secular evolution, our recovered state, used within an unscented Kalman filter (UKF), accurately reproduces the full trajectory, including its atmospheric reentry in August 2020, to within one day of the true decay time. This result demonstrates the viability of filtered TLE statistics as a proxy for precise orbit determination, particularly in dynamical regimes where resonances and long-period perturbations dominate. The techniques presented here provide a framework for trajectory reconstruction and prediction using only publicly available data and are broadly applicable to the study of high-altitude debris objects and legacy space missions whose original tracking and covariance data are unavailable.

\end{abstract}

\begin{keyword} 
Cislunar space;
Dynamical evolution and stability;
Gaussian mixture model; 
Space debris;
Space situational awareness;
Unscented Kalman filter
\end{keyword}

\end{frontmatter}

\parindent=0.5 cm

\section{Introduction}

The cislunar xGEO domain --- spanning from the outer bounds of geosynchronous orbit (GEO) to beyond the Moon \citep{kE59}  --- is rapidly emerging as a strategic and operationally vital region of space. As scientific, exploration, and commercial activity extend beyond traditional orbital regimes~\citep{mH21, aRdS21, bB24, aW25, dK25}, ensuring the long-term sustainability and predictability of objects in this domain becomes increasingly critical. Unlike low-Earth orbit (LEO) and GEO, where end-of-life disposal practices are well established through atmospheric reentry or graveyard orbits, no equivalent mitigation standards presently govern distant, high-eccentricity orbits (HEOs) that characterize much of the xGEO operational volume~\citep{cC15}. Often used for astrophysical, geophysical, and magnetospheric research~\citep{gL63, aB69, rK73, pB80, aB88, aG96, fJ01, nE03, nK14, gR15}, these orbits pose unique challenges to space situational awareness due to their complex resonance-driven dynamics and long-period secular perturbations from the Moon and Sun~\citep{bSjC66, gC67, zK67}. Orbits with ecliptic inclinations between approximately \(39.2^\circ\) and \(140.8^\circ\) are susceptible to von Zeipel-Lidov-Kozai (vZLK) oscillations, wherein the inclination and eccentricity undergo long-period coupled variations under the influence of a distant third body, while approximately preserving the semi-major axis~\citep{mR63, mA76, iS17, tI19, dA20}. These dynamical signatures are critical to understanding the evolution and predictability of xGEO objects~\citep{bL72}.

Although the fundamental dynamics of the cislunar environment are well established, accurately modeling and predicting the evolution of space objects in this regime remains a formidable challenge. The orbital motion of satellites in distant, high-eccentricity trajectories is often characterized by sensitivity to initial conditions and exposure to nonlinear dynamical regimes~\citep{aRbK25}, including resonance overlap and secular perturbations from the Moon and Sun. These effects can give rise to complex and, at times, chaotic orbital behavior, complicating both long-term forecasting and operational orbit determination. The difficulty is further compounded when tracking multiple uncooperative or inactive objects, for which maneuver activity and fragmentation events may remain unobserved or poorly constrained~\citep{nB21, pG23, aB25}. 

In contrast to the traditional geocentric domains, where orbit updates benefit from frequent tracking opportunities and near-periodic motion, satellites in the extended cislunar domain exhibit orbital periods on the order of several days to weeks~\citep{mA76, bE99}. This temporal sparsity of observation windows significantly limits the cadence of catalog maintenance and increases sensitivity to modeling errors in the intervening dynamical propagation~\citep{bL72}. Moreover, as objects traverse regions where traditional perturbations become dominant --- such as near lunar flybys or within resonance zones~\citep{rN59, sH62, rA63, mVkH14, cP24, aBdA24} --- standard assumptions of weak third-body effects no longer hold. These compounded uncertainties necessitate the development of refined filtering, propagation, and statistical-estimation techniques to ensure the fidelity of the space object catalog (SOC) in the xGEO and broader cislunar domain.

Historically, the dynamical sensitivity of cislunar orbits was first recognized in the early satellite era. Explorer VI and XII demonstrated that lunisolar perturbations could drastically alter orbital lifetime depending on launch timing~\citep{pM59, yKcW59, rS62, wHgM68}. Analytical work by Kozai, Musen, and others revealed that the secular evolution of the argument of perigee under solar and lunar influences can induce dramatic changes in perigee altitude~\citep{bSjC66, gC67, zK67, mR70, eR70, bL72, gJeR76, Liu1986, rN02}, leading either to premature decay or prolonged orbital residence. These insights shifted mission planning philosophy and introduced the concept of ``launch windows'' --- originally used for interplanetary trajectories --- to Earth-centric missions, recognizing that the initial configuration of the orbit relative to celestial geometry has profound long-term consequences~\citep{hM63, aM64, sPbS65, eR70}.

A striking example is the Orbiting Geophysical Observatory OGO-1 (1964-054A), colloquially referred to as EGO \citep{hM63}, which was launched into an orbit of semi-major axis 12.7 Earth radii and eccentricity $e \sim 0.918$. Early predictions projected atmospheric reentry by the early 1980s~\citep{bSjC66}, yet this object remained aloft until its eventual decay in August 2020. This discrepancy --- nearly four decades of predictive error --- highlights both the challenges in long-term orbit forecasting and the limitations of analytic models such as \texttt{SGP4} when applied to objects in xGEO-like regimes. Notably, EGO's launch occurred at a phase in its long-period vZLK cycle that led to an early and sustained perigee boost. A deviation of only ten minutes in launch timing would have dramatically altered its secular trajectory, likely resulting in reentry decades earlier~\citep{aW18,hN19}.

In light of renewed attention to the Earth-Moon system, there is a pressing need for tools that enable accurate trajectory recovery and prediction based solely on publicly available data \citep{dSmH24, dH25}. In this work, we present a statistically grounded methodology to reconstruct initial conditions from batches of historical two-line element sets (TLEs). The approach begins by evaluating TLEs with \texttt{SGP4} at their respective epochs to obtain osculating state vectors. These are then numerically propagated to a common epoch, transformed to orbital elements, and processed using fast-Fourier-transform-based numerical averaging to extract their mean-element representations~\citep{jS64, cU73, tE15}. To mitigate the influence of low-quality or inconsistent TLEs, we employ a Gaussian-mixture model (GMM) to statistically isolate a core subset of dynamically consistent mean elements. From this filtered population, a representative osculating element is reconstituted, yielding a Cartesian state vector that can be used as a high-fidelity initial condition. This methodology combines and extends the frameworks introduced by \citet{cLwM11} and \citet{hN19} by incorporating dynamical averaging and probabilistic filtering in mean-element space.

This recovered state not only serves as a dynamically meaningful reconstruction of the satellite's configuration at epoch, but also enables precise forward propagation. We demonstrate the utility of the method through application to the case of OGO-1. Despite the span of five decades and significant dynamical sensitivity, our recovered state predicts the satellite's final reentry date within one day of the true decay epoch. Moreover, we integrate the filtered initial condition into an unscented Kalman filter (UKF) to enable orbit estimation, state refinement, and covariance estimation. The methodology presented here provides a practical and statistically robust approach to long-arc orbit reconstruction and forecasting, with direct implications for xGEO situational awareness, catalog completeness, and long-term sustainability in the cislunar environment.

\section{The cislunar space-object catalog beyond GEO}

The publicly available catalog of Earth-orbiting objects, distributed in the form of TLEs, is maintained by the 18th Space Defense Squadron (18 SDS) of the United States Space Force. As the authoritative entity responsible for space object tracking on behalf of the U.S. Department of Defense, the 18 SDS curates the space-object catalog (SOC), which includes active satellites, inert upper stages, mission-related debris, and fragmentations. Public dissemination of TLEs is facilitated through the official data-sharing platform Space-Track.org, where registered users can access up-to-date and historical orbital elements for thousands of trackable objects in circumterrestrial orbits.

\begin{figure}[t!]
	\begin{center}
        \vskip -0.15in
	\includegraphics[width=0.925\textwidth]{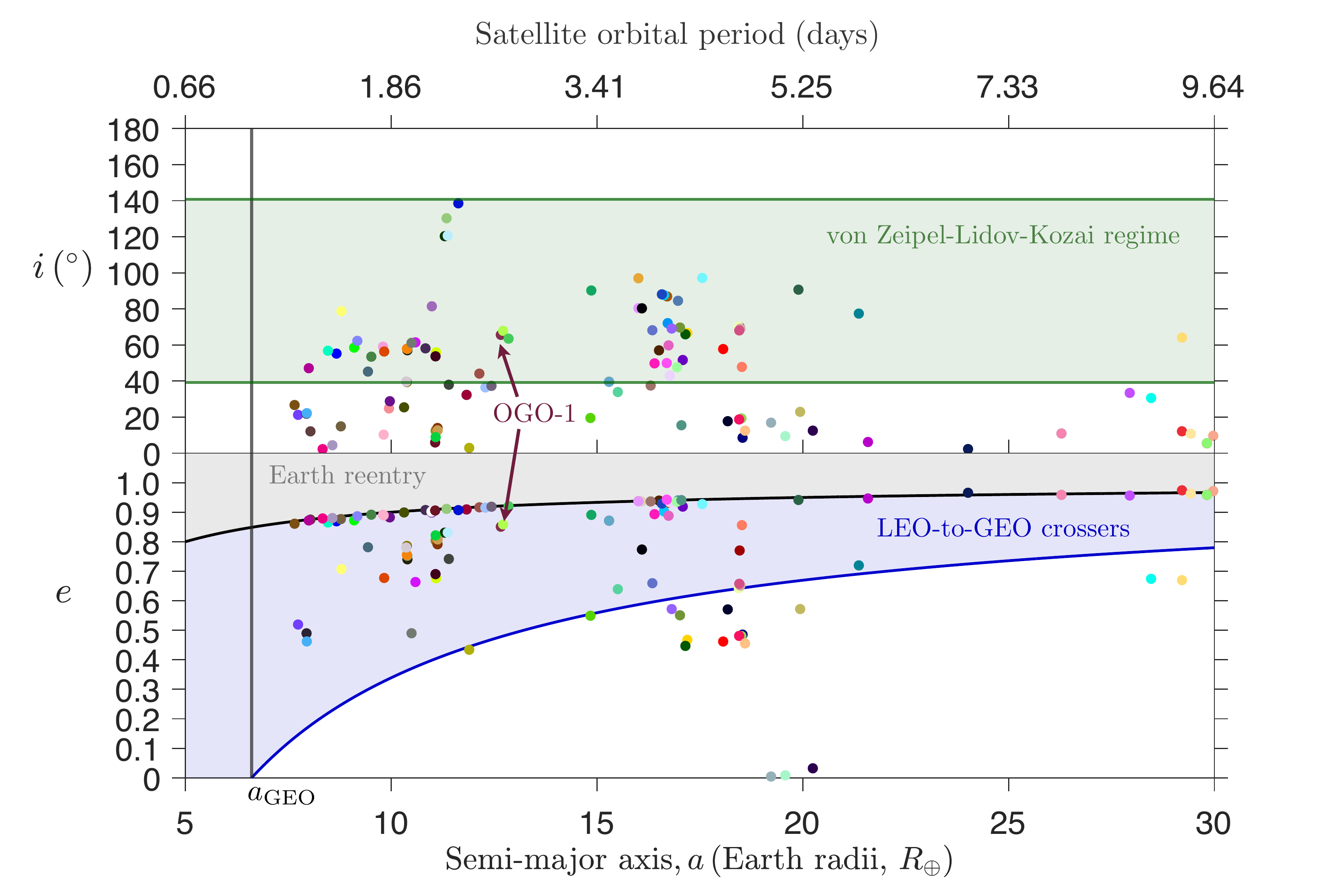}
	\vskip -0.05in
	\caption{\small 
 A snapshot of the historic and current cataloged xGEO space objects in the planes of semimajor axis-ecliptic inclination ($a,i$) ({\it top panel}) and semimajor axis-eccentricity ($a,e$) ({\it bottom panel}), where the {\it colored circles} correspond to the osculating elements obtained from the latest TLE of each object. The geocentric semi-major axis of GEO is indicated by the {\it vertical line} at $a$$\sim$$6.6 R_\E$. Objects that reach the Earth-grazing $a(1-e)=R_\E$ {\it curve} may reenter the atmosphere. The {\it light-blue shaded region} highlights LEO-to-GEO crossing orbits, where perigee falls below GEO while apogee extends beyond it --- a class of highly eccentric, dynamically unstable trajectories. The {\it upper green band} denotes the von Zeipel-Lidov-Kozai (vZLK) regime, defined by ecliptic inclinations in the range $\left( 39.2^\circ, 140.8^\circ \right)$, where secular lunisolar perturbations can induce large-amplitude oscillations in eccentricity and inclination over long timescales. NASA Orbiting Geophysical Observatory (OGO-1; satellite no. 1964-­054A) is indicated for reference. \\ \textit{Data source:} \url{www.space-track.org}. Assessed 4 May 2025.  
        }
	\label{fig:xGEO_snap}
	\end{center}
	\vskip -0.25in
	\hspace{2cm}\rule{12.5cm}{0.5pt}
	\vskip -0.15in
\end{figure}

While the majority of cataloged objects reside in LEO, medium-Earth orbit (MEO), or GEO, there exists a significant population of historical and operational spacecraft beyond the graveyard belt. Figure~\ref{fig:xGEO_snap} presents a snapshot of more than one hundred such objects in this cislunar regime  --- often referred to collectively as xGEO --- illustrating their distribution in semi-major axis-eccentricity and semi-major axis-inclination space. These objects, including legacy science missions, deep-space technology demonstrators, and long-lived debris, follow orbits that traverse xGEO space and interact dynamically with the Moon over multi-year or even multi-decade timescales~\citep{bSjC66, gC67}. Despite their sparse observational coverage and irregular maintenance, these distant objects are tracked and updated in the SOC, underscoring the necessity for enhanced dynamical models and data-filtering techniques to ensure catalog fidelity and situational awareness in the Earth-Moon environment.

Though widely utilized for decades, there remain persistent misconceptions surrounding the interpretation, accuracy, and inherent limitations of TLEs, as well as the analytic framework underpinning them --- the Simplified General Perturbations (\texttt{SGP4}) theory~\citep{fH04}. TLEs do not provide covariance information, and their positional accuracy at epoch varies significantly by regime: approximately several hundred meters to a few kilometers (1-$\sigma$) in LEO, a few kilometers in medium-Earth orbit (MEO), and tens of kilometers for GEO and HEOs~\citep{tF08, cFtS12, mHdA25}. Because TLEs are expressed in Brouwer-Lyddane mean elements, their associated osculating states must be reconstructed by reintroducing short-periodic variations through \texttt{SGP4} evaluation at the native epoch.

\begin{figure}[b!]
	\begin{center}
        \vskip -0.025in
	\includegraphics[width=0.975\textwidth]{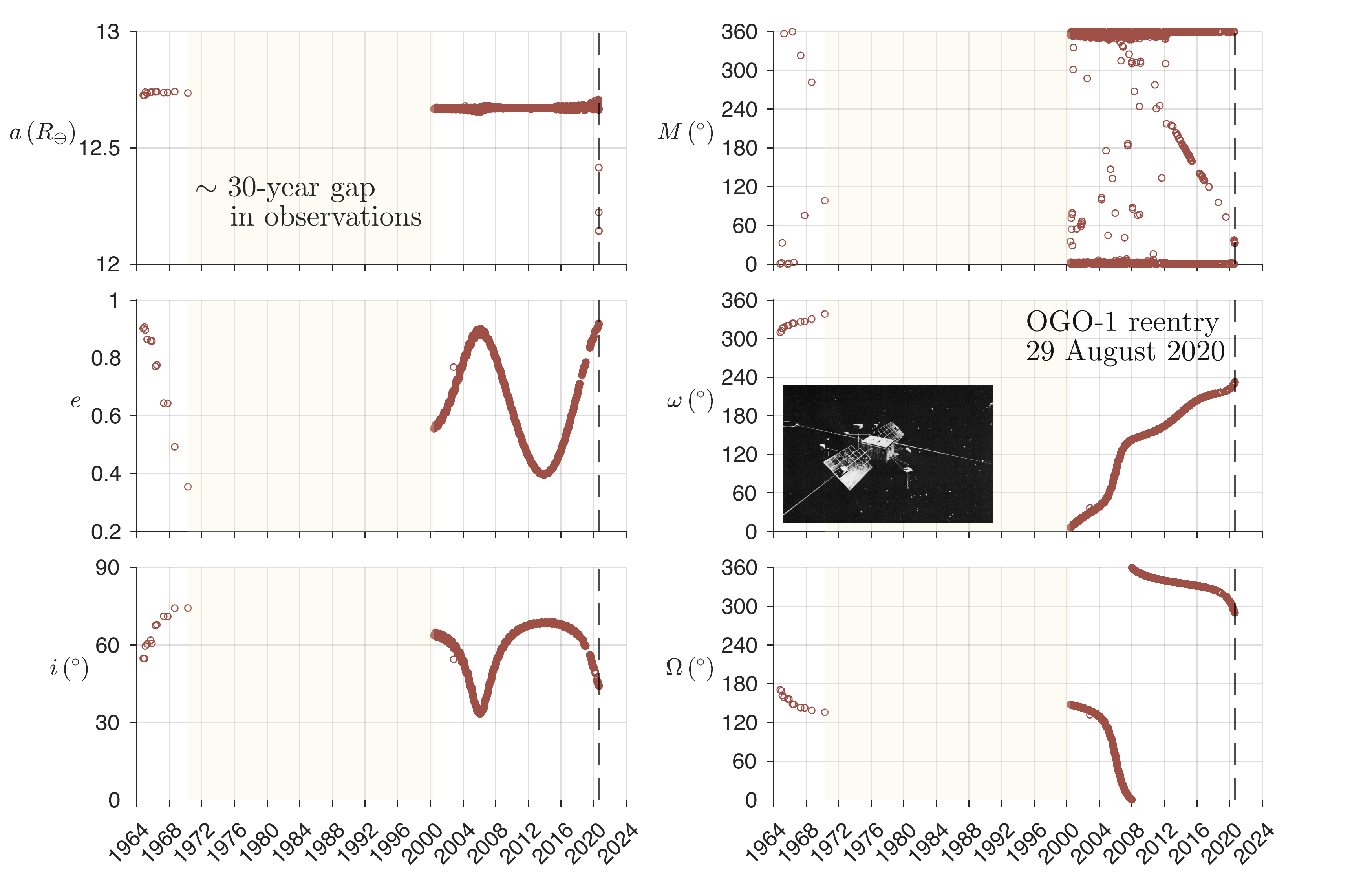}
	\vskip -0.05in
	\caption{\small 
Time history of the osculating Keplerian elements of NASA's Orbiting Geophysical Observatory 1 (OGO 1, 1964-054A), derived from TLEs using the \texttt{SGP4} algorithm at each native epoch. The orbital elements are measured with respect to the ecliptic frame. This xGEO satellite was launched on 5 September 1964 01:23 UTC (MJD 38643.05764)  into a distant, highly eccentric orbit ($a \sim 12.7\,R_\oplus$, $e \sim 0.918$), exhibiting pronounced secular variation in inclination and argument of perigee due to lunar perturbations. Tracking was discontinued for over three decades, as evidenced by the absence of publicly available TLEs between approximately 1971 and 2001. OGO 1 reentered Earth's atmosphere on 29 August 2020 20:44 UTC (MJD 59090.86389), nearly 56 years after launch. The difficulty of maintaining long-arc predictions for such orbits underscores the need for improved filtering and initialization methods in cislunar space. \\ \textit{Data source:} \url{www.space-track.org}. Assessed 4 May 2025.
        }
	\label{fig:OGO_history}
	\end{center}
	\vskip -0.25in
	\hspace{2cm}\rule{12.5cm}{0.75pt}
	\vskip -0.05in
\end{figure}

Figure~\ref{fig:OGO_history} presents the full multi-decadal time history of the osculating orbital elements of OGO-1, reconstructed from historical TLEs using \texttt{SGP4} evaluation at their respective epochs. All elements are referenced in the ecliptic frame. The top-left panel shows the semi-major axis, which remains relatively constant near $12.7\,R_\oplus$ from the mid 2000s through the satellite's final decay, consistent with the expected secular invariance of $a$. The eccentricity and inclination panels (left column, middle and bottom) display prominent long-period variations, indicative of resonant (i.e., vZLK) and third-body perturbations acting over the course of decades. Notably, the argument of perigee (middle-right) undergoes smooth secular evolution, while the mean anomaly (top-right) and longitude of ascending node (bottom-right) exhibit significant discontinuities and wrapping effects due to observational gaps.

A shaded region in the background denotes a roughly 30-year interval --- spanning from the early 1970s until the mid 2000s --- during which OGO-1 was absent from the SOC. During this period, the object was effectively untracked, perhaps reflecting the challenges of maintaining custody of high-eccentricity, slow-moving objects in xGEO space. A dashed vertical line marks the date of OGO-1's eventual atmospheric reentry on 29 August 2020. The dense cluster of observations preceding this date illustrates the renewed tracking effort in the final decades of the satellite's life. Taken together, the figure underscores the long-term dynamical evolution of distant orbits, the limitations of historical observation practices, and the need for improved archival continuity in cislunar xGEO catalog maintenance.

\section{Coherency-gated filtering philosophy and robust state reconstruction}

To obtain a higher-fidelity initial condition (IC) from historical TLE data, a number of studies have treated the \texttt{SGP4}-reconstructed osculating states as pseudo-observations and applied batch least-squares differential correction using high-precision numerical propagators~\citep{cLwM11, jS13, cCfH18, dL20, mHdA25}. This approach has demonstrated substantial improvements in predictive accuracy over propagation with \texttt{SGP4} alone. In particular, \citet{cLwM11} showed that by fitting an orbit to a TLE window and propagating it with a high-fidelity dynamical model, one may achieve an order-of-magnitude improvement in short- to medium-term orbit prediction. 

Several subsequent efforts have sought to improve this reconstruction process by first identifying and discarding anomalous TLEs prior to the orbit-fitting process. \citet{aL19}, for instance, proposed a straightforward temporal consistency check based on residual thresholds between propagated TLEs and observed ephemerides to reject outliers, while \citet{jL21} introduced a statistical filter on residuals computed from batch-corrected TLEs to suppress corrupted entries. More recently, \citet{dSmH24} explored a heuristic anomaly score derived from element-wise deviations and dynamic inconsistencies across a TLE batch. Separately, \citet{jL21} proposed an expectation-maximization (EM)-based outlier detection method to improve TLE time-series filtering, while \citet{dSmH24} applied particle filtering for online anomaly detection in TLEs with a particular focus on maneuver detection.

While these data-driven approaches have yielded notable improvements, their pre-filtering criteria rely primarily on geometric residuals or statistical heuristics, without explicit consideration of the underlying dynamical structure. As a result, they may struggle in regimes where short-term discrepancies do not adequately reflect deeper dynamical inconsistencies, such as those arising from resonant interactions or secular drift. In contrast, the methodology presented here employs a statistically principled and dynamically coherent framework for isolating a subset of mean elements consistent with the long-term secular evolution of the orbit. By filtering in the mean, rather than osculating, element space and leveraging statistical clustering to define coherency, our method offers a principled and dynamics-consistent mechanism for TLE selection prior to reconstruction. This distinction is especially critical in resonance-dominated or high-eccentricity regions, where short-term residuals can be misleading and dynamical consistency must be judged over longer timescales.

\begin{algorithm}[t!]
\caption{\scriptsize Coherency-Gated Filtering and Initial Condition Reconstruction from TLE Ensembles}
\label{alg:TLE_GMM_Filtering}
\scriptsize
\begin{algorithmic}[1]
\Require TLE batch $\mathcal{T} = \{T_i\}_{i=1}^N$, target epoch $t_0$, propagator $\mathcal{P}$, arc mode $\mathsf{mode} \in \{\text{short}, \text{long}\}$, element set $\mathcal{E}$
\Ensure Robust Cartesian state $\bm{x}(t_0)$ at reference epoch $t_0$
\vspace{1mm}
\Statex \textbf{Step 1: Evaluate \texttt{SGP4} at each TLE's native epoch}
\For{$T_i \in \mathcal{T}$}
    \State Compute Cartesian state $\bm{x}_i(t_i) = \texttt{SGP4}(T_i)$
\EndFor
\State Define set $\mathcal{X}(t_i) = \{ \bm{x}_i(t_i) \}_{i=1}^N$
\vspace{1mm}
\Statex \textbf{Step 2: Propagate all states to common epoch $t_0$}
\For{$\bm{x}_i(t_i) \in \mathcal{X}(t_i)$}
    \State Propagate to $t_0$: $\bm{x}_i(t_0) = \mathcal{P}(t_i \rightarrow t_0; \bm{x}_i)$
\EndFor
\State Assemble $\mathcal{X}(t_0) = \{ \bm{x}_i(t_0) \}_{i=1}^N$
\vspace{1mm}
\Statex \textbf{Step 3: Convert propagated states to osculating elements at $t_0$}
\For{$\bm{x}_i(t_0) \in \mathcal{X}(t_0)$}
    \State Convert: $\textbf{\oe}_i(t_0) = \mathrm{cart2osc}(\bm{x}_i(t_0))$
\EndFor
\State Form osculating-elements matrix $\textbf{\oe}^{\mathrm{osc}} = [\textbf{\oe}_1(t_0), \ldots, \textbf{\oe}_N(t_0)]^\top$
\vspace{1mm}
\Statex \textbf{Step 4: Compute mean elements via numerical averaging}
\For{$\textbf{\oe}_i(t_0) \in \textbf{\oe}^{\mathrm{osc}}$}
    \State Generate short arc about $t_0$; compute mean $\overline{\textbf{\oe}}_i$ via FFT-based averaging
\EndFor
\State Construct mean-elements matrix $\overline{\textbf{\oe}} = [\overline{\textbf{\oe}}_1, \ldots, \overline{\textbf{\oe}}_N]^\top$
\vspace{1mm}
\Statex \textbf{Step 5: Apply element-wise outlier detection}
\For{$k = 1$ to $|\mathcal{E}|$}
    \If{$\mathcal{E}_k = \text{MA}$}
        \State De-weight by setting GMM mode to `percentile' or `adaptive' with low threshold (e.g., 15th percentile)
    \EndIf
    \If{$\mathsf{mode} = \text{short}$}
        \State Apply MAD-based outlier detection to column $k$ of $\overline{\textbf{\oe}}$
    \Else
        \State Attempt GMM-based filtering on column $k$ of $\overline{\textbf{\oe}}$
        \If{GMM fails or flags all}
            \State Apply fallback method (e.g., MAD, quantile, or z-score)
        \EndIf
    \EndIf
    \State Store inlier mask $\bm{m}_k$ for element $k$
\EndFor
\State Combine inliers: $\bm{m}_{\text{all}} = \bigwedge_{k=1}^{|\mathcal{E}|} \bm{m}_k$
\vspace{1mm}
\Statex \textbf{Step 6: Reconstruct $\textbf{\oe}^*(t_0)$ from osculating inliers at $t_0$}
\State Let $\mathcal{I} = \{ i \mid \bm{m}_{\text{all}}(i) = 1 \}$
\If{$|\mathcal{I}| = 0$}
    \State Use $\textbf{\oe}^*(t_0) = \mathrm{median}(\textbf{\oe}^{\mathrm{osc}})$
\ElsIf{$|\mathcal{I}| = 1$}
    \State Use the corresponding state: $\textbf{\oe}^*(t_0) = \textbf{\oe}_i(t_0)$
\ElsIf{$|\mathcal{I}| \leq 3$}
    \State Use element-wise median of $\{ \textbf{\oe}_i(t_0) \}_{i \in \mathcal{I}}$
\Else
    \State Use element-wise mean of $\{ \textbf{\oe}_i(t_0) \}_{i \in \mathcal{I}}$
\EndIf
\vspace{1mm}
\Statex \textbf{Step 7: Convert consensus elements to Cartesian state}
\State Convert: $\bm{x}(t_0) = \mathrm{osc2cart}(\textbf{\oe}(t_0))$
\vspace{1mm}
\State \Return $\textbf{\oe}(t_0)$ and $\bm{x}(t_0)$
\end{algorithmic}
\end{algorithm}

\begin{figure}[t!]
\centering
\resizebox{0.6\textwidth}{!}{%
\begin{tikzpicture}[node distance=0.33cm and 1.65cm, every node/.style={align=center},
  block/.style = {rectangle, draw, fill=blue!5, text width=9.6em, font=\scriptsize, text centered, rounded corners, minimum height=2.4em},
  process/.style = {rectangle, draw, fill=green!10, text width=9.6em, font=\scriptsize, text centered, rounded corners, minimum height=2.4em},
  iot/.style = {trapezium, trapezium left angle=70, trapezium right angle=110, draw, fill=orange!10, minimum height=2.4em, text width=6.5em, font=\scriptsize, text centered},
  iob/.style = {trapezium, trapezium left angle=70, trapezium right angle=110, draw, fill=orange!10, minimum height=2.4em, text width=10em, font=\scriptsize, text centered},  
  decision/.style = {diamond, draw, fill=gray!10, text width=5.8em, font=\scriptsize, text badly centered, inner sep=1pt},
  line/.style = {draw, thick, -latex'}
  ]

\node [iot] (input) {Batch of TLEs};
\node [block, below=of input] (sgp4) {\texttt{SGP4} at $t_i$ \\ to get $\bm{x}_i(t_i)$};
\node [block, below=of sgp4] (prop) {Propagate $\bm{x}_i(t_i)$ \\ to $t_0$ via $\mathcal{P}$};
\node [block, below=of prop] (elconv) {Convert $\bm{x}_i(t_0) \rightarrow \textbf{\oe}_i(t_0)$};
\node [block, below=of elconv] (avg) {Numerical averaging \\ $\textbf{\oe}_i \rightarrow \overline{\textbf{\oe}}_i$};
\node [block, below=of avg] (stack) {Form matrix $\overline{\textbf{\oe}}$};
\node [decision, below=of stack] (modebranch) {Arc mode?};
\node [block, below=of modebranch] (madgmm) {Apply MAD or GMM \\ + fallback on $\overline{\textbf{\oe}}$};
\node [block, below=of madgmm] (inlierextract) {Extract osculating \\ inliers $\mathcal{I}$};
\node [block, below=of inlierextract] (inlieravg) {Element-wise medium or mean \\ of $\textbf{\oe}_i(t_0) \in \mathcal{I}$};
\node [block, below=of inlieravg] (finalcart) {Convert $\textbf{\oe}(t_0) \rightarrow \bm{x}_0$};

\node [process, right=1.8cm of stack] (ukfinit) {Initialize UKF \\ with $\bm{x}_0$, $\bm{P}_0$};
\node [process, below=of ukfinit] (sigpts) {Generate \\ sigma points};
\node [process, below=of sigpts] (proppts) {Propagate through \\ $\mathcal{F}$};
\node [process, below=of proppts] (measmap) {Map through \\ $\mathcal{H}$};
\node [process, below=of measmap] (kalman) {Compute $\hat{\bm{x}}, \bm{K}$, \\ update $\bm{x}, \bm{P}$};
\node [iob, below=of kalman] (output) {State estimate \\ $\bm{x}(t_k)$};

\path [line] (input) -- (sgp4);
\path [line] (sgp4) -- (prop);
\path [line] (prop) -- (elconv);
\path [line] (elconv) -- (avg);
\path [line] (avg) -- (stack);
\path [line] (stack) -- (modebranch);
\path [line] (modebranch) -- (madgmm);
\path [line] (madgmm) -- (inlierextract);
\path [line] (inlierextract) -- (inlieravg);
\path [line] (inlieravg) -- (finalcart);
\path [line] (finalcart.east) -- ++(0.75,0) |- (ukfinit);
\path [line] (ukfinit) -- (sigpts);
\path [line] (sigpts) -- (proppts);
\path [line] (proppts) -- (measmap);
\path [line] (measmap) -- (kalman);
\path [line] (kalman) -- (output);

\end{tikzpicture}%
}
\caption{Schematic of the coherency-gated initialization process (\textit{left}) for robust TLE-based state estimation. The mean elements are filtered using arc-dependent strategies (MAD or GMM with fallback), while the final state is computed using osculating inliers. This estimate seeds a UKF update sequence (\textit{right}).}
\label{fig:UKF_GMM_flow}
\end{figure}
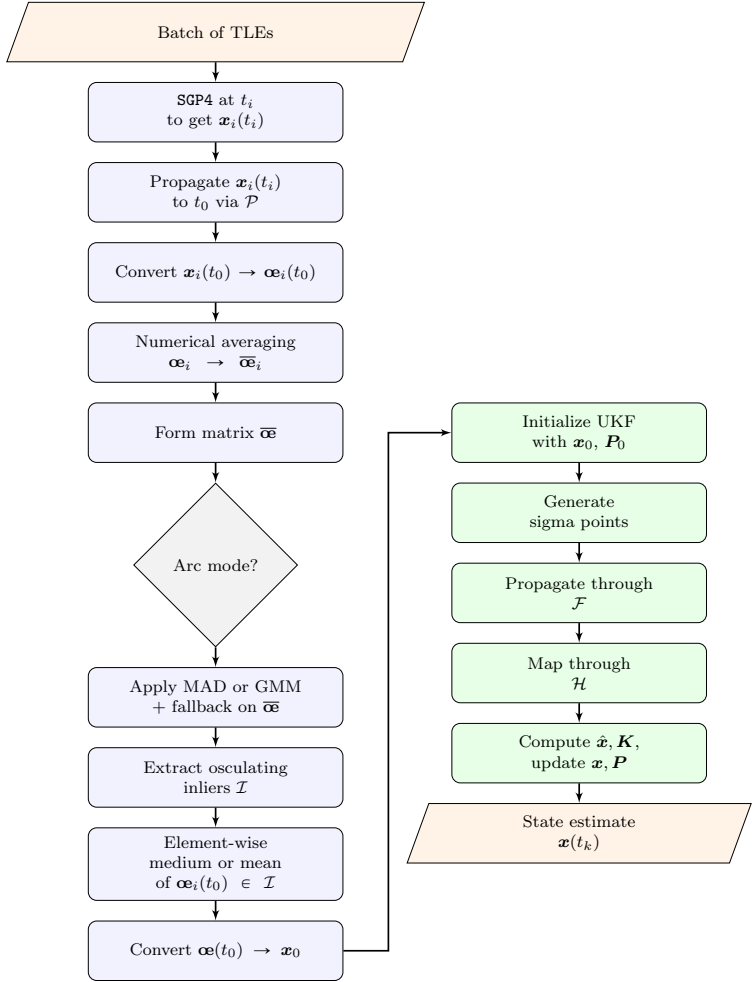

Building on these foundational efforts, we develop a complementary approach rooted in ensemble statistical filtering of historical TLEs with perturbation-theoretic averaging, outlined in Algorithm~\ref{alg:TLE_GMM_Filtering} and shown schematically in Figure~\ref{fig:UKF_GMM_flow} (left). To reconstruct a high-fidelity IC from a batch of TLEs, we introduce a coherency-gated filtering framework that systematically identifies and excludes inconsistent members of the ensemble. The central premise is that while TLEs represent the same physical object, they may differ significantly in fidelity due to differing fit epochs, propagation errors, or systemic biases introduced during generation. For each TLE, we first apply the \texttt{SGP4} model at its native epoch to obtain the corresponding Cartesian state vector, $\bm{x}_i (t_i) = [\bm{r}_i, \bm{v}_i]^T$. These state vectors are then numerically propagated to a common epoch $t_0$ using a high-fidelity orbit propagator, producing a set $\{\bm{x}_i (t_0)\}$. At this epoch, each state is converted to osculating orbital elements, denoted $\textbf{\oe}_i (t_0)$, from which we extract the corresponding mean elements $\overline{\textbf{\oe}}_i$ via fast Fourier transform (FFT)-based numerical averaging, following the techniques introduced by \citet{jS64} \citep[see, also,][]{cU73, tE15}. The averaging process removes high-frequency variations and permits a direct comparison of secular trends among the TLEs. 

Outlier detection is performed element-wise using either robust statistical estimators (e.g., median absolute deviation (MAD)) or probabilistic classification via Gaussian mixture models (GMMs), with fallback logic triggered adaptively depending on arc span and data morphology. For long arcs, GMM-based classification with adaptive thresholds can better accommodate secular divergence while preserving coherence. For short arcs --- where physical drift is minimal and scatter is dominated by measurement error --- MAD is preferred for its robustness. In cases of particularly noisy or complex data, or when processing very large datasets comprising thousands of TLEs for a single object, GMM and MAD can be applied in tandem; with GMM first identifying the dominant cores of the distribution, and MAD subsequently refining the inlier set. The intersection of all inlier sets defines a subset of TLEs whose secular orbital content is statistically consistent. We then compute the average of the inlier osculating elements, which serves as the recovered initial condition. This process yields a state that is dynamically coherent, minimally biased, and insensitive to noisy or corrupted TLEs within the batch.

Beyond its use in reconstructing a high-fidelity initial condition at a fixed epoch, this algorithm is further employed to estimate spacecraft states at intermediate epochs between the native spans of the TLEs. Given a window of TLEs whose epochs are non-uniformly distributed in time, we extract all \texttt{SGP4}-evaluated state vectors $\{\bm{x}_i (t_i)\}$, propagate them to a common intermediate epoch $t_j$, and apply the same numerical averaging and GMM- or MAD-based probabilistic filtering procedure described previously. This enables the generation of statistically robust state estimates even at epochs not directly represented in the original TLE data. By treating the batch of temporally local TLEs as a stochastic ensemble, this approach mitigates the effects of catalog sparsity and measurement irregularity in orbit-determination frameworks such as batched weighted least squares or sequential Bayesian estimators (e.g., the UKF). Applying the method over a sliding or adaptive temporal window allows for continuous reconstruction of secular-consistent spacecraft states, providing a means to bridge gaps between TLE epochs with dynamically filtered interpolants (Algorithm~\ref{alg:UKF_Coherency_Gate}). In doing so, it enables enhanced accuracy in trajectory estimation, ephemeris generation, and residual analysis, particularly for defunct or untracked space objects where direct measurement updates are unavailable.

\begin{algorithm}[t!]
\caption{\scriptsize UKF-Based Orbit Estimation with Coherency-Gated Initial Conditions from TLE Ensembles}
\label{alg:UKF_Coherency_Gate}
\scriptsize
\begin{algorithmic}[1]
\Require Windowed TLE batch $\mathcal{T} = \{T_i\}_{i=1}^N$, target epoch $t_0$, dynamics model $\mathcal{F}$, measurement model $\mathcal{H}$, process noise $\bm{Q}$, measurement noise $\bm{R}$, measurement sequence $\{\bm{z}_k\}$
\Ensure State estimate $\bm{x}(t_k)$ and covariance $\bm{P}(t_k)$ at each measurement epoch

\Statex \textbf{Initialization via Coherency-Gated TLE Ensemble}
\For{$T_i \in \mathcal{T}$}
    \State Evaluate \texttt{SGP4} at TLE epoch $t_i$ to obtain $\bm{x}_i(t_i)$
    \State Propagate $\bm{x}_i(t_i)$ to $t_0$ via high-fidelity propagator $\mathcal{P}$ \hfill\texttt{Cowell-type (e.g., \texttt{ASSIST})}
    \State Convert $\bm{x}_i(t_0) \rightarrow \textbf{\oe}_i(t_0)$ and apply FFT-based averaging: $\textbf{\oe}_i(t_0) \rightarrow \overline{\textbf{\oe}}_i$
\EndFor
\State Form matrix $\overline{\textbf{\oe}} = [\overline{\textbf{\oe}}_1, \ldots, \overline{\textbf{\oe}}_N]^\top$
\State Apply arc-dependent outlier filtering (MAD or GMM with fallback) to $\overline{\textbf{\oe}}$ to identify inliers $\mathbf{m}_{\text{all}}$
\State Extract corresponding osculating inliers $\textbf{\oe}_i(t_0)$ for inlier set $\mathcal{I}$
\State Reconstruct $\textbf{\oe}^{*}(t_0)$ using element-wise median or mean over $\{\textbf{\oe}_i(t_0)\}_{i \in \mathcal{I}}$
\State Convert $\textbf{\oe}^{*}(t_0) \rightarrow \bm{x}_0$
\State Initialize UKF: $\bm{x}(t_0) \gets \bm{x}_0$, $\bm{P}(t_0) \gets \bm{P}_0$ (user-defined or empirical)

\Statex \textbf{Unscented Kalman Filter Propagation and Update}
\For{each measurement time $t_k$ after $t_j$}
    \State Generate sigma points from $\bm{x}(t_{k-1})$ and $\bm{P}(t_{k-1})$
    \For{each sigma point $\chi_j$}
        \State Propagate $\chi_j$ to $t_k$ using dynamics model $\mathcal{F}$
    \EndFor
    \State Compute predicted mean $\hat{\bm{x}}^{-}(t_k)$ and covariance $\bm{P}^{-}(t_k)$
    \State Transform propagated sigma points through measurement model $\mathcal{H}$
    \State Compute predicted measurement $\hat{\bm{z}}(t_k)$ and innovation covariance $\bm{S}$
    \State Compute cross-covariance $\bm{C}_{xz}$ and Kalman gain $\bm{K}$
    \State Update state: $\bm{x}(t_k) = \hat{\bm{x}}^{-}(t_k) + \bm{K}[\bm{z}_k - \hat{\bm{z}}(t_k)]$
    \State Update covariance: $\bm{P}(t_k) = \bm{P}^{-}(t_k) - \bm{K}\bm{S}\bm{K}^T$
\EndFor

\State \Return $\bm{x}(t_k), \bm{P}(t_k)$ for all $t_k$
\end{algorithmic}
\end{algorithm}

In the following section, we demonstrate the application of this coherency-gated filtering approach to a challenging historical case: the long-arc TLE record of OGO-1, whose extreme eccentricity, vZLK oscillations, and poorly constrained tracking history provide an ideal testbed for robust IC recovery.  

\section{Reconstructing the orbit of OGO-1}

The non-negligible collision risk posed by LEO-to-xGEO transiting spacecraft motivates both theoretical study and practical implementation of filtering techniques. Using ephemeris-quality orbit propagators, such as the \texttt{IAS15} integrator \citep{hR19} available within the \texttt{REBOUND} and \texttt{ASSIST} software packages \citep{mHetal23}, we can accurately handle orbital evolution that is highly sensitive to initial conditions, including close approaches and resonances. These modern Solar-System dynamics integrators were validated against NASA's \texttt{GMAT} \citep{sH14} for the propagation of cislunar xGEO objects.

\begin{figure}[!t]
	\begin{center}
        \vskip -0.025in
	\includegraphics[width=0.925\textwidth]{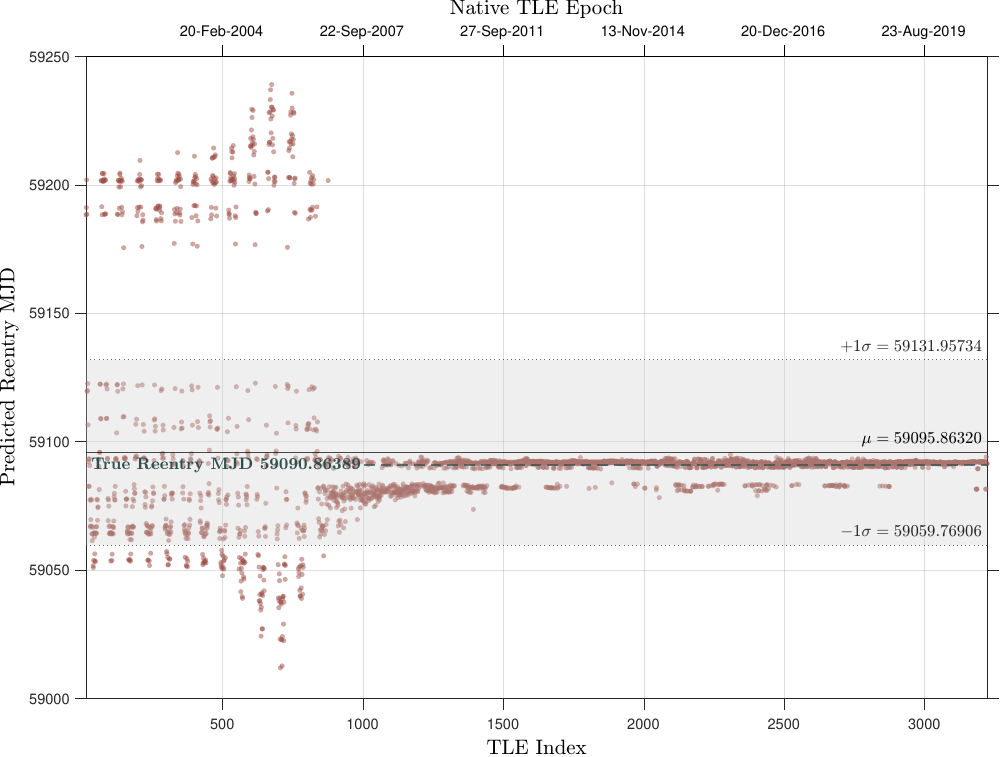}
	\vskip -0.05in
	\caption{\small 
Reentry epoch predictions for OGO-1, derived from forward propagation of individual TLEs under high-fidelity dynamics. The distribution reveals substantial variance in predicted decay dates, a manifestation of both observational sparsity and structural inconsistencies in the raw TLE record. The dense cluster near the true reentry date contrasts with broad outliers and two prominent clusters in the post-2007 data, underscoring the challenges of recovering a coherent dynamical state in the absence of ensemble filtering.
        }
	\label{fig:OGO_reentry}
	\end{center}
	\vskip -0.25in
	\hspace{2cm}\rule{12.5cm}{0.75pt}
	\vskip -0.05in
\end{figure}

Figure~\ref{fig:OGO_reentry} shows the predicted reentry epochs obtained by propagating each individual TLE of OGO-1 to decay, with no ensemble filtering or coherency gating applied. Each TLE was converted to its osculating state at epoch and integrated forward under high-fidelity dynamics. The resulting wide dispersion of predicted reentry dates --- spanning nearly a full year --- reflects the extreme sensitivity of long-term predictions to errors and inconsistencies inherent in raw TLE data. Although each TLE nominally represents the same physical object, differences in fit quality, epoch regression, and dynamical coherence can produce substantial divergence in the resulting orbital forecast. Notably, the reentry map exhibits two distinct clusterings of predicted decay epochs in the post-2007 data (beyond TLE index $\sim$1000), which likely reflect systematic shifts in TLE generation procedures or sensor coverage across this historical record.\footnote{Here, ``reentry'' is defined as the osculating perigee falling below 50 km altitude --- the threshold adopted in \citet{cC15} for Earth-grazing HEO analyses, chosen specifically to guarantee that the atmospheric drag impulse in a single perigee pass irreversibly captures the spacecraft and prevents it from re-emerging.}

Moreover, given the sparse cadence of available TLEs for OGO-1 (roughly one resolution every 2--3 orbits) and the large variability between individual TLE-derived predictions evidenced in the reentry map, the batch recovery method of \citet{cLwM11}, or modern adaptations, is unlikely to succeed in yielding a corrected initial condition. Their approach hinges upon fitting an orbit across a semi-coherent window of TLEs and leveraging a high-fidelity dynamical model to refine prediction; however, in the case of OGO-1, the inherent noise and inconsistencies within the TLE data themselves, exacerbated by the wide observation gaps, produces discordant orbital content across the batch. The intervening intervals of several orbits introduce phase ambiguities, while the statistical scatter within the TLEs prevents the filter from resolving a consistent secular trend. As a result, any state recovered through batch or Kalman filtering remains ill-posed and fails to achieve improved predictive performance, even when propagated with an accurate dynamical model. This highlights the necessity of ensemble-based filtering to isolate a dynamically consistent subset prior to initial condition generation and trajectory reconstruction.

In the present work, the identification of outliers is performed not on the osculating elements ${\textbf{\oe}_i^{\mathrm{osc}}(t_0)}$, but rather on the set of corresponding mean elements ${\overline{\textbf{\oe}}_i(t_0)}$. This distinction is fundamental to the philosophy of the coherency gate. Each TLE, although ostensibly representing the same physical object, encodes a pseudo-observation derived through the nonlinear reconstruction of orbital states from sparse measurements. In the process of TLE generation --- specifically the \texttt{SGP4}-compatible mapping of observational data into Brouwer-Lyddane mean elements and the subsequent regression to the last ascending node --- error is introduced not merely as measurement scatter, but as structural ambiguities. This is especially pronounced in the mean anomaly, whose value is frequently decorrelated from any physically meaningful time origin. It is thus common for TLE epochs themselves to become detached from the underlying dynamical behavior they purport to represent.

\begin{figure}[t!]
	\begin{center}
        \vskip -0.025in
	\includegraphics[width=0.75\textwidth]{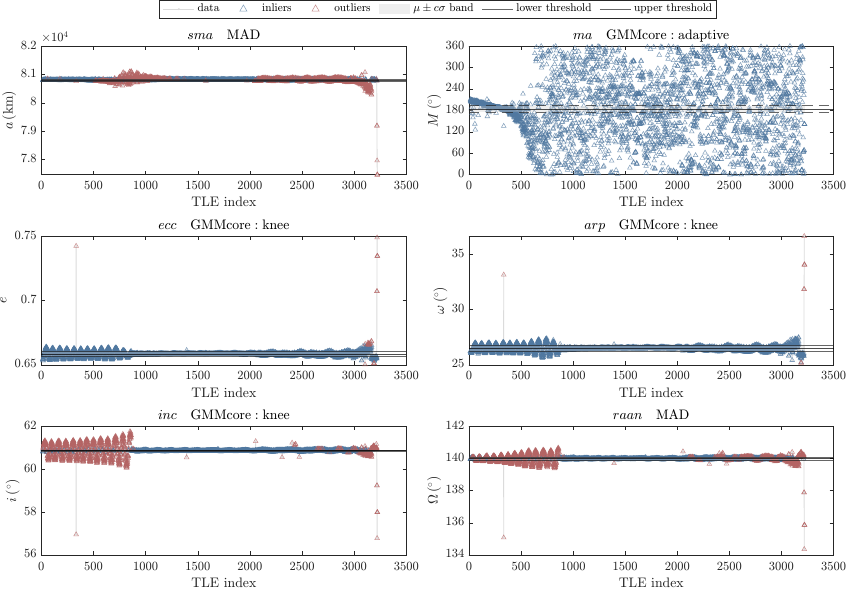}
	\includegraphics[width=0.75\textwidth]{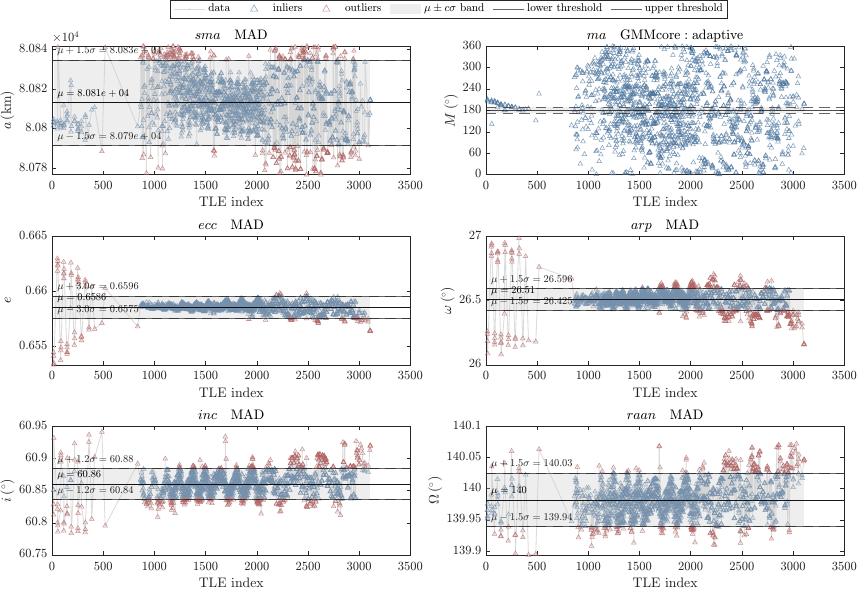}
	\vskip -0.05in
	\caption{\small 
Application of Algorithm~\ref{alg:TLE_GMM_Filtering} to the approximately 3,200 TLEs of OGO-1 spanning from 23 June 2000 to 20 August 2020, with the coherency gate epoch set to 6 July 2002. A Gaussian mixture model (GMM) with fallback logic was first applied ({\it top panel}), followed by a median absolute deviation (MAD) filter ({\it bottom panel}). The element-wise inlier mean elements are depicted as {\it blue triangles}, while detected outliers are shown in {\it red-orange}. For each orbital element, this tandem filtering approach --- combining GMM core detection and MAD refinement --- yields a consensus osculating $\textbf{\oe}^*(t_0)$ at the selected epoch. This state represents the most dynamically coherent and statistically faithful reconstruction of the satellite's orbit, derived from the ensemble of inlier osculating elements.
        }
	\label{fig:OGO_filter}
	\end{center}
	\vskip -0.25in
	\hspace{2cm}\rule{12.5cm}{0.75pt}
	\vskip -0.05in
\end{figure}
 
\subsection{Recovered state and long-arc trajectory prediction}

To illustrate the practical application of this filtering philosophy, we apply the coherency-gated framework of Algorithm~\ref{alg:TLE_GMM_Filtering} to the long-arc historical TLE dataset of the OGO-1 spacecraft, a canonical example of a high-eccentricity, resonance-perturbed xGEO object. The recovered osculating elements at the coherency gate epoch, along with the predicted reentry date obtained from \texttt{ASSIST} propagation of each consensus state, are summarized in Table~\ref{tab:OGO_IC}. Each initial condition was generated by independent filter runs at epoch 6 July 2002 7:34 UTC (MJD 52461.31528), with slight differences arising from the filter mode and weighting. Notably, both runs predict the reentry date of OGO-1 to within 5 hours of the actual decay on 29 August 2020 20:44 UTC (MJD 59090.86389), despite a prediction arc spanning nearly two decades.

\begin{table}[h!]
\small
\centering
\caption{Recovered osculating elements at coherency gate epoch (MJD 52461.31528) and predicted reentry date for OGO-1 (1964-054A).}
\label{tab:OGO_IC}
\begin{tabular}{lccccccc}
\toprule
Predicted Reentry [MJD] & $a$ [km] & $e$ & $i$ [deg] & $\Omega$ [deg] & $\omega$ [deg] & $M$ [deg] \\
\midrule
59090.75562 & 80810.22906 & 0.65582 & 60.84363 & 139.90280 & 26.16947 & 193.22310 \\
59090.67808 &  80809.59825 & 0.65588 & 60.84355 & 139.90139 & 26.17832 & 197.37701 \\
\bottomrule
\end{tabular}
\end{table}

\begin{figure}[t!]
	\begin{center}
        \vskip -0.025in
	\includegraphics[width=1.0\textwidth]{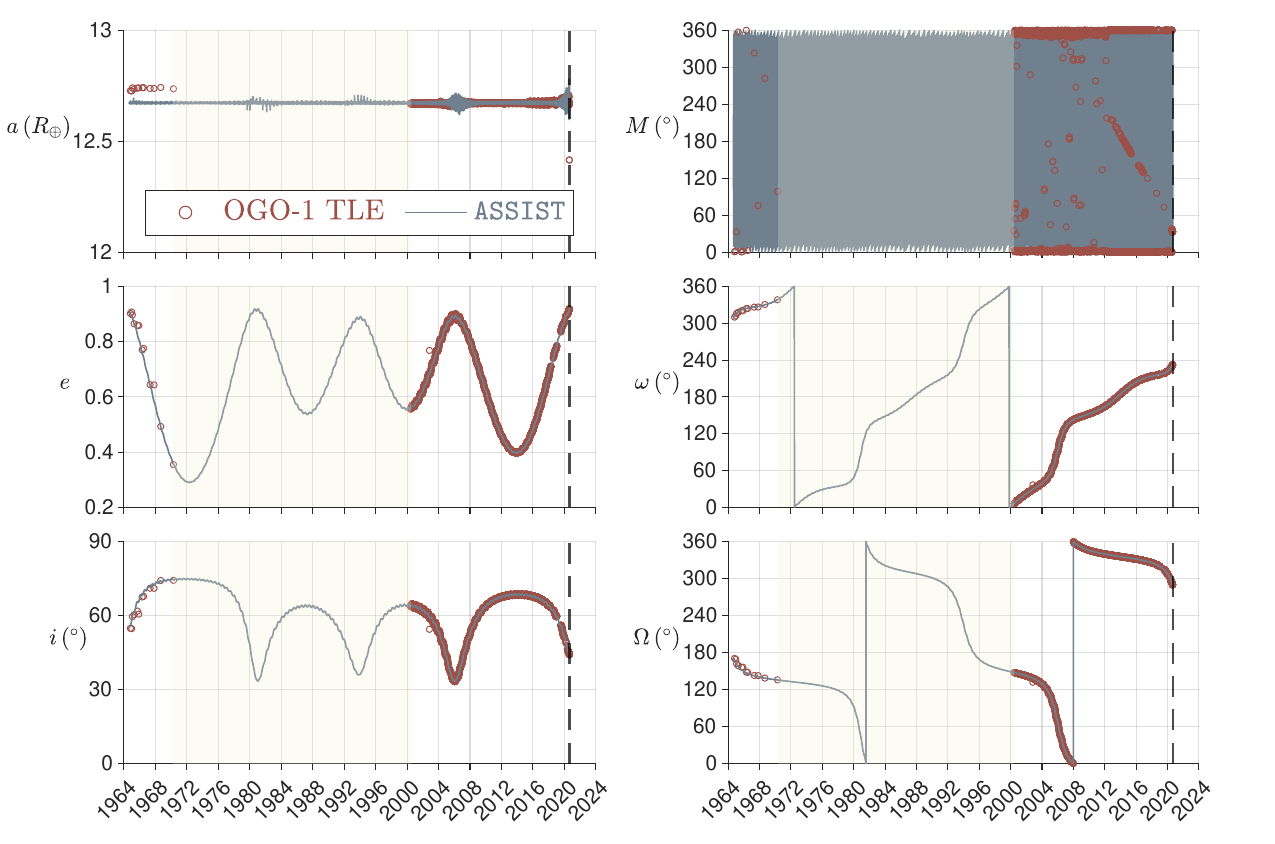}
	\vskip -0.05in
	\caption{\small 
The multi-decadal orbital evolution of NASA's Orbiting Geophysical Observatory 1 (OGO 1, 1964-054A), obtained using an \texttt{ASSIST} prediction from the recovered state at the coherency gate epoch of 6 July 2002 7:34 UTC (MJD 52461.31528), compared to the historical TLE time history. All angular elements are referenced to the ecliptic plane. The reconstructed state yields a predicted reentry epoch within 5 hours of the true atmospheric decay on 29 August 2020 20:44 UTC (MJD 59090.86389). The secular evolution of semi-major axis, eccentricity, inclination, argument of perigee, and right ascension of the ascending node are all well captured, reflecting the long-period vZLK-driven oscillations that dominated OGO-1’s final decades in orbit. 
        }
	\label{fig:OGO_recovery}
	\end{center}
	\vskip -0.25in
	\hspace{2cm}\rule{12.5cm}{0.75pt}
	\vskip -0.05in
\end{figure}

Figure~\ref{fig:OGO_recovery} presents the resulting long-arc trajectory reconstruction, showing the evolution of OGO-1's orbital elements under \texttt{ASSIST} propagation from the recovered initial condition, compared to the full TLE time history. The secular evolution of the semi-major axis, eccentricity, inclination, argument of perigee, and node are all well captured, including the signature of the long-period vZLK oscillations that modulated the satellite's perigee over the decades. The excellent agreement underscores both the predictive utility of the recovered initial condition and the effectiveness of the coherency-gated filtering process in isolating dynamically meaningful states from noisy historical data.

\clearpage 

\subsection{Application of UKF filtering for sparse TLE records}

\begin{figure}[!t]
	\begin{center}
        \vskip -0.025in
	\includegraphics[width=0.925\textwidth]{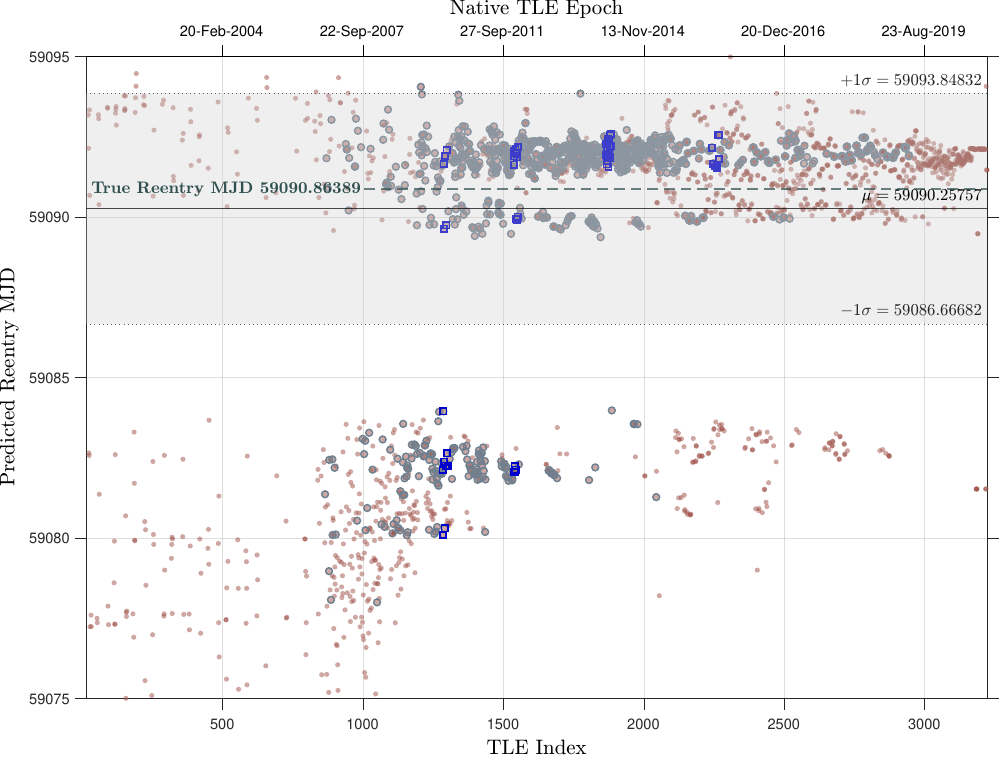}
	\vskip -0.05in
	\caption{\small 
Reentry epoch predictions for OGO-1, showing the ensemble of raw TLE-based predictions (background, as in Figure~\ref{fig:OGO_reentry}), the recovered inlier set ({\it gray circles}, from Figure~\ref{fig:OGO_filter}, {\it bottom panel}), and the four temporally localized TLE windows ({\it blue squares}) selected for UKF-based state recovery. The displayed $\mu \pm \sigma$ band reflects the statistics of the inlier set, not the broader dispersion shown in Figure~\ref{fig:OGO_reentry}. These windows (centered near 2010, 2012, 2014, and 2016) provide the stochastic ensembles used to estimate filtered spacecraft states at intermediate epochs.
        }
	\label{fig:OGO_inliers}
	\end{center}
	\vskip -0.25in
	\hspace{2cm}\rule{12.5cm}{0.75pt}
	\vskip -0.05in
\end{figure}

Figure~\ref{fig:OGO_inliers} illustrates the background context for the UKF-based orbit estimation applied to OGO-1. The methodology, summarized in Figure~\ref{fig:UKF_GMM_flow} and detailed in Algorithm~\ref{alg:UKF_Coherency_Gate}, is applied independently to four selected TLE windows (shown as blue squares in Figure~\ref{fig:OGO_inliers}), centered near 2010, 2012, 2014, and 2016. For each window, an ensemble of temporally localized TLEs is first coherency-gated to extract a subset of dynamically consistent inliers, following the robust filtering procedure described earlier. To generate state estimates at intermediate epochs --- including times between the actual TLE indices --- the full ensemble of TLEs within each window is propagated to a series of target epochs, forming a synthetic measurement sequence. This enables continuous UKF-based estimation even at epochs not directly sampled by the native TLE record. The overall approach mitigates the effects of TLE sparsity and epoch irregularity by providing statistically coherent initializations and synthetic measurements for each windowed UKF sequence. The resulting osculating inlier sets then serve as the initialization for the unscented Kalman filter (UKF), enabling propagation of filtered state estimates and covariances to intermediate epochs. 

Table~\ref{tab:OGO_UKF} summarizes the recovered osculating elements and predicted reentry dates obtained by applying the UKF-based orbit estimation independently to each of the four selected windows. While we do not expect this UKF application to outperform the secular-filtered trajectory presented earlier (Figure~\ref{fig:OGO_recovery}, Table~\ref{tab:OGO_IC}), the purpose here is to demonstrate the viability of applying the coherency-gated UKF methodology over shorter timescales, using sparse TLE data. In this case, the selected windows are relatively close to the final reentry epoch and span comparatively short arcs, emphasizing the generation of local initial conditions and associated covariances suitable for dynamic state estimation. The UKF runs, seeded by the windowed inlier ensembles, successfully converge to coherent states and yield predicted reentry dates that remain consistent across the four windows. This illustrates how the methodology can provide robust filtered estimates even when applied to sparse, irregular TLE datasets; offering a practical tool for enhancing orbit estimation frameworks where long arcs or secular fits are not feasible.

\begin{table}[H]
\small
\centering
\caption{Recovered osculating elements at the end of each UKF-selected window and predicted reentry date for OGO-1 (1964-054A).}
\label{tab:OGO_UKF}
\begin{tabular}{llccccccc}
\toprule
Predicted Reentry & Epoch [MJD] & $a$ [km] & $e$ & $i$ [deg] & $\Omega$ [deg] & $\omega$ [deg] & $M$ [deg] \\
\midrule
59092.584447 & 55246.20907 & 80839.92 & 0.60906 & 64.60 & 345.24 & 152.937 & 296.55 \\
59092.485288 & 55985.04247 & 80823.63 & 0.45177 & 67.98 & 339.52 & 165.82 & 355.39 \\
59092.130638 & 56707.64540 & 80814.94 & 0.40104 & 68.82 & 335.52 & 186.22 & 359.67 \\
59091.865304 & 57451.38049 & 80813.17 & 0.49624 & 67.89 & 331.41 & 205.21 & 0.93 \\
\bottomrule
\end{tabular}
\end{table}

\section{Conclusion}

We have presented a dynamical coherency-gating framework for recovering physically consistent initial conditions from historical TLEs, enabling accurate reconstruction of long-arc trajectories for distant and highly eccentric Earth satellites. Mean elements --- derived through integration rather than inference --- serve as the statistical foundation in which true object-to-object variance reflects dynamical coherence or incompatibility rather than artifact. By using \(\overline{\textbf{\oe}}_i(t_0)\) as our filtering substrate, we explicitly acknowledge the constructed and often ill-posed nature of raw TLEs, restoring epistemic clarity by isolating those epochs whose encoded states remain consistent with the consensus geometry of the underlying object.

This work demonstrates that properly filtered TLE statistics can be leveraged for high-accuracy orbit recovery even in dynamically sensitive regions such as cislunar xGEO space, where traditional \texttt{SGP4}-based predictions degrade over time. The methodology is broadly applicable to long-lived space debris and legacy missions in xGEO, addressing critical gaps in current space situational awareness and orbit determination capabilities as interest in the Earth-Moon system continues to grow. Beyond the retrospective, this work functions as a critique of persistent community-wide negligence in orbital design for long-term sustainability. The failure to anticipate or even account for the secular consequences of high-eccentricity orbits has produced debris legacies that defy short-horizon planning. Our framework provides both a tool for reconstructive dynamics and a call for a more nuanced understanding of the lasting imprint of orbital choices.

Crucially, the coherency-gating approach can be integrated directly into operational catalogs and conjunction-analysis pipelines. By flagging and down-weighting inconsistent TLEs in real time, satellite operators and debris-monitoring agencies can maintain a dynamically coherent object database without resorting to expensive tasking or precision tracking assets for every object. This capability not only improves forecast accuracy in regions of complex perturbations, but also enables targeted re-observation campaigns where the reconstructed uncertainties exceed mission-critical thresholds.

Finally, as humanity's footprint extends beyond Earth and into cislunar space, the need for an informed orbital stewardship ethos becomes ever more pressing. The insights gained here about the long-term impact of perigee oscillations, resonance interactions, and secular drift should inform both future mission design and regulatory guidelines. By combining robust statistical filtering with principled dynamical modeling, we offer a path toward sustainable orbital regimes in xGEO, one where historical knowledge, operational practice, and policy evolve in concert to preserve the space environment for future generations.  

\section*{Acknowledgements}

A.R. acknowledges support by the Air Force Office of Scientific Research (AFOSR) under Grant No. FA9550-24-1-0194. 

\bibliographystyle{elsarticle-harv}
\bibliography{asr_tle}

\end{document}